# Maximizable informational entropy as measure of probabilistic uncertainty

C.J. Ou, A. El Kaabouchi, L. Nivanen, F. Tsobnang,

A. Le Méhauté, Q.A. Wang

Institut Supérieur des Matériaux et Mécaniques Avancés du Mans, 44 Av. Bartholdi,
72000 Le Mans, France

**Abstract**

In this work, we consider a recently proposed entropy $I$ (called varentropy) defined by a variational relationship $dI = \beta(\overline{dx} - \overline{dx})$ as a measure of uncertainty of random variable $x$. By definition, varentropy underlies a generalized virtual work principle $\overline{dx} = 0$ leading to maximum entropy $d(I - \beta\overline{x}) = 0$. This paper presents an analytical investigation of this maximizable entropy for several distributions such as stretched exponential distribution, $\kappa$-exponential distribution and Cauchy distribution.

## 1) Introduction

The term 'information' has been used as a measure of uncertainty of probability distribution in information theory[1][2]. In this sense, it is often called entropy with the detriment to possible confusion with the entropy of the second law of thermodynamics which has another specific physical definition for equilibrium system. In the present study of the relationship between uncertainty and probability, we would like to use the term 'varentropy' in order to distinguish the present uncertainty measure from both 'information' and 'entropy' for a reason we will explain below.

The search for the functional relationship between entropy and associated probability distribution has since long been a question in statistical and informational science. There are many relationships directly postulated or established on the basis of the presumed properties of entropy. The reader can refer to the references [1] to [10] to see several examples of entropies proposed on the basis of postulated entropy properties. Among the numbers of proposed entropies, the Shannon informational entropy $I = -\sum_i p_i \ln p_i$ is singled out as unique maximizable formula of uncertainty measure as claimed by Jaynes in its maximum entropy principle (maxent) [12][13]. Nevertheless, it is actually believed that other formula can be useful in some physics contexts. There are already many statistical theories constructed on the maximization of entropies which are postulated or deduced from assumed properties of entropy[6-10]. The correctness of these entropies is then verified through the validity of corresponding probability distributions often derived by using maxent.

The extension of maxent raises several questions about the validity and limit of that logic. The first question is why entropy, or uncertainty measure, can be maximized for assigning probability distribution. The second question is which entropies, among the numbers of entropies formula proposed up to now, each measuring a kind of probabilistic uncertainty, can be maximized, and why? Within the maxent of Jaynes, these questions are answered with unique Shannon entropy advocated by considering the anthropologic interpretation of subjective probability. However, these arguments are at odds with the frequency interpretation of probability widely accepted by physicists. Another questions relevant to the above ones is, given the variety of probability distribution observed in Nature, what is the right uncertainty measures which can yield them at maximum?

A possible way of answering these questions is to invert the logic of maxent, i.e., instead of postulating entropy property and formula and then maximizing them with additional principles and hypothesis concerning the constraints (of which the underlying reasons are often ambiguous so that certain constraints are somewhat arbitrary) to get useful probability, we only define a generic measure. This methodology was tested in [11] with a variational definition $dI = \beta(d\bar{x} - \overline{dx})$ of a measure $I$ of uncertainty for the simple situation of one random variable $x$, where $\beta$ is a characteristic constant. $I$ is called *varentropy* due to its variational definition implying a maximum state satisfying an extended Lagrange-d'Alembert variational principle of virtual work $\overline{dx} = 0$ [14]. Varentropy is different from the usual notion of information in that it is defined without any assumption about its physical signification (e.g., missing information or ignorance), its property (additive or not) and functional form. It is also different from the thermodynamic entropy of second law because it is not necessarily related to heat. Nevertheless, it can be the second law entropy in a special case of equilibrium thermodynamic system.

In this work, we investigate further the varentropy for three well known probability distributions: the stretched exponential distribution, the κ-exponential and the Cauchy distribution. The first is often observed in complex relaxation process of nonequilibrium system [15]. The second is a probability distribution proposed in the case of relativistic complex system [7]. The third one is often used to describe the line shape in spectroscopy [16].

## 2) Varentropy for stretched exponential distribution

According to its definition, with the hypothesis of complete normalization $\sum_i p_i = 1$ and the expectation $\bar{x} = \sum_i p_i x_i$, the varentropy can be given by (let $\beta = 1$).

$$dI = d\bar{x} - \overline{dx} = \sum_i x_i dp_i, \qquad (1)$$

where $\{x_i, i = 1,2,3\cdots\}$ represent the possible value of random variable $x$ at different state with corresponding probability $\{p_i, i = 1,2,3\cdots\}$. The stretched probability distribution is given by

$$p_i = \frac{1}{Z}\exp\left(-x_i^\alpha\right) \text{ with } \alpha > 0, \qquad (2)$$

where $Z$ is a normalization constant or sometimes called the partition function. From Eq.(2) we can easily get

$$x_i = \left(\ln\left(\frac{1}{p_i Z}\right)\right)^{\frac{1}{\alpha}}. \tag{3}$$

Substituting Eq.(3) into Eq.(1) we can get

$$dI = \sum_i \left(\ln\left(\frac{1}{p_i Z}\right)\right)^{\frac{1}{\alpha}} dp_i. \tag{4}$$

Then the varentropy can be written as

$$I = \int dI = \sum_i \int \left(\ln\left(\frac{1}{p_i Z}\right)\right)^{\frac{1}{\alpha}} dp_i = \frac{1}{Z}\sum_i \Gamma\left(\frac{1}{\alpha}+1, -\ln(p_i Z)\right) + C, \tag{5}$$

where $\Gamma(a,z) \equiv \int_z^\infty t^{a-1}\exp(-t)dt$ is the incomplete gamma function which will reduce to complete gamma function at the limit $z \mapsto 0$, and $C$ is the integral constant. In order to determine $C$ we consider a system without uncertainty as a special case of Eq.(5). Suppose the system has only two states $i=1$ and 2 with $p_1 =1$ and $p_2 =0$, the uncertainty of the system is zero requires

$$0 = \frac{1}{Z}\Gamma\left(\frac{1}{\alpha}+1, -\ln Z\right) + \frac{1}{Z}\Gamma\left(\frac{1}{\alpha}+1, -\ln 0\right) + C, \tag{6}$$

From definition of incomplete gamma function, the second item in the right hand side of Eq.(6) tend to zero. Now from Eq. (6) we can get the integral constant

$$C = -\frac{1}{Z}\Gamma\left(\frac{1}{\alpha}+1, -\ln Z\right). \tag{7}$$

Substituting Eq.(7) into Eq.(5) yields

$$I = \frac{1}{Z}\sum_i \Gamma\left(\frac{1}{\alpha}+1, -\ln(p_i Z)\right) - \frac{1}{Z}\Gamma\left(\frac{1}{\alpha}+1, -\ln Z\right). \tag{8}$$

This is the varentropy for the stretched exponential distribution. In fact if we choose a variable replacement as: $x_i \mapsto a + bx_i^{'}$ where the parameters $a$ and $b$ are associated with the Lagrange multipliers, then the normalization constant $Z = \sum_i \exp(-x_i^\alpha)$ can be replaced by $Z^{'} = \sum_i \exp\left(-(a+bx_i^{'})^\alpha\right) = 1$. So Eq.(8) can be rewritten as

$$I' = \sum_i \Gamma\left(\frac{1}{\alpha}+1,-\ln(p_i')\right) - \Gamma\left(\frac{1}{\alpha}+1\right), \tag{9}$$

with the corresponding distribution function

$$p_i' = \exp\left(-(a+bx_i')^\alpha\right). \tag{10}$$

Eqs.(9) and (10) are in accordance with the results in Ref.[18] while the derivation given here seems more straightforward.

### 3) Varentropy for κ-exponential distribution

With the help of the definition of κ-exponential function [19]

$$\exp_{\{\kappa\}}(x) = \left(\kappa x + \sqrt{1+\kappa^2 x^2}\right)^{1/\kappa}, \tag{11}$$

the κ-exponential distribution is given by

$$p_i = \frac{1}{Z}\exp_{\{\kappa\}}\left(-\beta(E_i-\mu)\right), \tag{12}$$

where $\kappa$ is a deformation parameter. $E_i$ and $\mu$ represent the energy of the system at the $i$-th microstate and the chemical potential of the system, respectively. It's obviously that Eq.(12) will reduce to the Maxwell-Boltzmann distribution at the $\kappa \to 0$ limit. One easily get the inverse function of $p_i$ as

$$E_i = -\frac{1}{\beta}\left(\frac{(p_i Z)^\kappa - (p_i Z)^{-\kappa}}{2\kappa}\right) + \mu. \tag{13}$$

Since $\{E_i\}$ represent the energy of each microstate of the system, it's can be considered as a random variable. Then from Eq. (1) the information measurement of such a system can be written as

$$dI = \beta \sum_i E_i dp_i. \tag{14}$$

where $\beta$ is a constant with inverse dimension of energy. Then $\beta E_i$ is a dimensionless variable. Substituting Eq. (13) into Eq. (14) yields

$$dI = \sum_i \left(-\left(\frac{(p_i Z)^\kappa - (p_i Z)^{-\kappa}}{2\kappa}\right) + \beta\mu\right) dp_i. \tag{15}$$

The varentropy of the κ-exponential distribution directly reads,

$$I = \sum_i \left( c(\kappa) p_i^{1+\kappa} + c(-\kappa) p_i^{1-\kappa} \right) + C, \qquad (16)$$

where

$$c(\kappa) = -\frac{Z^\kappa}{2\kappa(1+\kappa)}. \qquad (17)$$

In the same way we can determine the integral constant $C = -(c(\kappa) + c(-\kappa))$. These results appearing in accordance with the one in Ref. [19] was not by chance. In fact the definition of Eq. (1) is very general and by this method we can get even more different measurements of uncertainty (sometimes it was called entropic forms by other authors) based on the different observed probability distributions in the nature.

In the above calculation, the random variable is energy as in the original version of $\kappa$-statistics. The result is however valid for any random variable.

## 4) Varentropy for Cauchy distribution

The Cauchy distribution function is given by

$$p_i = \frac{1}{\pi\gamma \left( 1 + \left( \frac{x_i - x_0}{\gamma} \right)^2 \right)}, \qquad (18)$$

where $x_0$ is a constant which specifies the location of the peak of Cauchy distribution, and $\gamma$ is the half-width of the line shape of distribution at half-maximum. For the sack of convenience, we can choose a standard Cauchy distribution at $x_0 = 0$, so Eq.(18) reads

$$p_i = \frac{1}{Z} \frac{1}{\left(1 + (x_i/\gamma)^2\right)}, \qquad (19)$$

where $Z$ is a normalization constant. If $x_i$ is a continuous random variable in the region $]-\infty, \infty[$, the normalization constant $Z$ will be equal to $\pi$ after a simple integral. From Eq.(19) we can get the definition of mean value of $\{x_i\}$ which does not exist in the original definition of Cauchy distribution,

$$\bar{x} = \frac{1}{Z} \int_{-A}^{A} \frac{x}{1 + (x/\gamma)^2} dx, \qquad (20)$$

with

$$Z = \int_{-A}^{A} \frac{1}{1+(x/\gamma)^2} dx, \tag{21}$$

where $A$ is the maximum of random variable $x_i$. We can easily write

$$x_i = \pm\gamma\sqrt{\frac{1}{p_i Z}-1}. \quad \pm x_i = \gamma\sqrt{\frac{1}{p_i Z}-1} \tag{22}$$

Substituting Eq. (22) into the definition of varentropy, i.e., Eq.(1), after integration, one gets

$$I = \frac{\gamma}{Z}\sum_i \left(\frac{\pm x_i}{1+x_i^2} \pm \arctan(x_i)\right) + C \tag{23}$$

$$= \frac{\gamma}{Z}\sum_i \left(p_i Z\sqrt{\frac{1}{p_i Z}-1} - \arctan\left(\sqrt{\frac{1}{p_i Z}-1}\right)\right) + C$$

where

$$C = \frac{\gamma}{Z}\left(\arctan(\sqrt{\frac{1}{Z}-1}) - Z\sqrt{\frac{1}{Z}-1} + \frac{\pi}{2}\right). \tag{24}$$

From Eqs. (23) and (24) we can get the curve of uncertainty measurement $I$ with respect of the probability distribution. As shown in Fig.1, for a two states system the $I$ first increased with the increasing of probability and then decreased. At $p_i = 0$ and $p_i = 1$ the uncertainty measurement are zero since the system was absolutely determined at one state and there has no uncertainty. So there exist maximum values for $I$ with different $\gamma$, it means that the expressions of Eqs. (23) and (24) can be maximized and consequently one will get the Cauchy distribution if the Maxent method was adopted with the constraints of normalization condition and the definition of expectation of random variables, i.e. $\bar{x} = \sum_i p_i x_i$.

## 5) Concluding remarks

The work in the present paper is an extension of the work in Ref. [11]. Based on the definition of uncertainty measurement for random variable, i.e. Eq. (1), we can derive several different entropic forms corresponding to different probability distribution functions. It's very easy to verify that all these entropies can be maximized with the constraints of mean value of random variable $\{x_i\}$ and the normalization condition of probability distribution. All these will result to the corresponding distributions which have been observed in the nature. It's worth to point out that the calculations in this paper are not inverse process of Maxent. First, there was not any concrete functional form required in the definition of uncertainty measurement, it means that Eq. (1) is very general and can be use to derive many different entropic forms only based on the existed distributions but no other additional assumptions. It can give reasonable explanations for some entropic forms such as stretched exponential one [9] which have been used without any introduction. Second, we prefer "uncertainty measurement" rather than "entropy" which may be confused with the one of thermodynamics. If and only if Maxwell-Boltzmann distribution was adopted, the corresponding uncertainty measurement reduced to Shannon entropy, which was firstly maximized by Jaynes. And Eq. (1) now has more concrete physical meaning; it's nothing but the first law of thermodynamics.

Acknowledgements

This work was supported by the Region des Pays de la Loire of France under grant number 2007-6088.

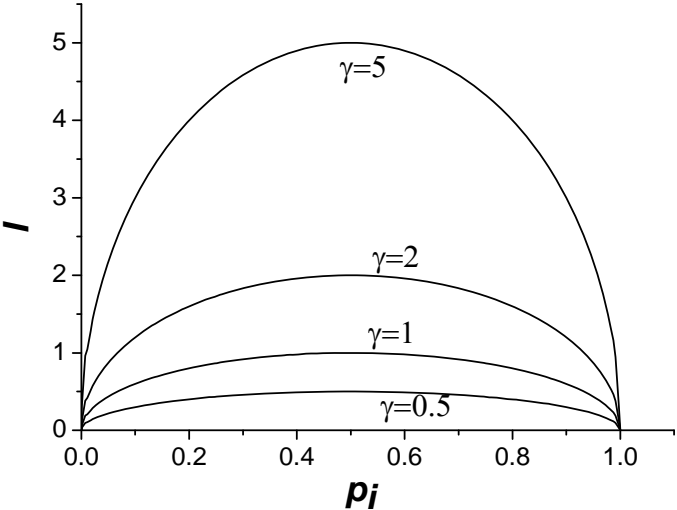

Fig 1. The variation of Cauchy varentropy versus probability distribution for a two states system and for different values of $\gamma$.